\documentclass[aps,prx,twocolumn,superscriptaddress,amsmath,amssymb,floatfix]{revtex4-2}

\usepackage{graphicx}
\usepackage{dcolumn}
\usepackage{bm}
\usepackage{xcolor}

\def\e{\epsilon}
\def\d{\dagger}
\def\pd{\phantom{\dagger}}

\def\k{\vec{k}}

\begin{document}


\title{Flat-Band Driven Kondo Breakdown and Reentrant Effects in Heavy-Fermion Moir\'{e} Superlattices
}


\author{Fabian Eickhoff}
 \affiliation{Institute of Software Technology, German Aerospace Center, 51147 Cologne, Germany}
 \email{fabian.eickhoff@dlr.de}
  
\author{Jian-Xin Zhu}%
 \affiliation{Theoretical Division, Los Alamos National Laboratory, Los Alamos, New Mexico 87545, USA}
 \affiliation{Center for Integrated Nanotechnologies, Los Alamos National Laboratory, Los Alamos, New Mexico 87545, USA}
 \email{jxzhu@lanl.gov}

\author{Benedikt Fauseweh}
 \affiliation{Institute of Software Technology, German Aerospace Center, 51147 Cologne, Germany}
 \affiliation{Department of Physics, Condensed Matter Theory, TU Dortmund University, 44221 Dortmund, Germany}
 \email{benedikt.fauseweh@tu-dortmund.de}

\date{\today}

\begin{abstract}
Moir\'{e} superlattices (MSL) in van der Waals heterostructures have demonstrated their incredible power in driving emergent electronic phenomena, some of which are reminiscent of those usually only observed in bulk strongly correlated quantum materials. With the recent discovery of van der Waals $f$-electron materials, the design of novel MSLs of intrinsic strong correlation is now within the reach. Here we study the novel electron phases of two-dimensional heavy-fermion MSL with increasingly diluted $f$-electron local moments. By applying  dynamical mean field theory with numerical renormalization group as an impurity solver, we demonstrate the appearance of a new energy scale and a re-entrant Kondo breakdown in connection with the emergence of a flat band in the system. We further compare our numerical findings with predictions derived from the Lieb-Mattis theorem and show the necessity of the new energy scale to consistently reconcile the predictions with the conventional single-impurity limit for exceedingly large unit cells.
\end{abstract}


\maketitle

\section{Introduction}

The study of heavy fermion materials has captivated researchers for decades due to their rich and complex physical properties, which arise from the interplay between localized \( f \)-electron states and itinerant conduction electrons \cite{HF}. These materials exhibit a variety of intriguing phenomena, including unconventional superconductivity, quantum criticality, and non-Fermi liquid behavior \cite{HF_superconductivity, HF_NFL, NFL, Doniach77, LQCP_3, LQCP_1}.  
Central to this complexity is the Kondo effect \cite{TK, Wilson}, where a magnetic impurity in a metal is screened by conduction electrons at the Kondo temperature \( T_\text{K} \), setting the scale for universality in thermodynamic and transport properties. In the heavy fermion community, the two energy scales,
 \( T^\text{loc}_0 \) and \( T^\text{glob}_0 \), and their relationship are heavily discussed.\cite{PAM_1, PAM_2} These scales set the onset of single-site Kondo screening at \( T = T^\text{loc}_0 \approx T_\text{K} \) and the onset of lattice coherence at \( T = T^\text{glob}_0 \).

So far,  these novel properties have been tuned mainly by such conventional nonthermal parameters as pressure, chemical doping, and magnetic field, in heavy fermion systems, in which the density of states of a wide conduction electron band is mostly featureless at the Fermi energy~\cite{AHewson1993}. 
In recent years, the Moir\'{e} superlattice (MSL) in graphene and other van der Waals (vdW) materials, has triggered tremendous theoretical and experimental interest, where novel electronic properties such as the Hofstadter Butterfly~\cite{LAPonomarenko2013,CRDean2013,BHunt2013,GLYu2014,LWang2015}, Brown-Zak oscillations~\cite{LAPonomarenko2013,CRDean2013,BHunt2013,GLYu2014,LWang2015,RKumar2017}, superconductivity~\cite{YCao2018a,MYankowitz2018,GChen2019}, and Mott-like insulating state~\cite{MYankowitz2018,YCao2018b,LWang2020} have been achieved. In these systems, the competition between the electron kinetic energy and the Coulomb repulsion is tuned by a significant increase of the real space length scale by twisting 2D vdW materials. We note that MSLs can arise from either construction of periodic patterns in the mismatched two sublattices due to twisting or lattice mismatch, or periodic deposition of adatoms with scanning tunneling microscopy~\cite{RJCelotta2014}.  Recent prediction~\cite{BGJang2022} and experimental realization~\cite{VAPosey2024} of the intrinsic vdW heavy-fermion systems as well as synthetic Moir\'{e} Kondo lattices~\cite{Zhao2023,PhysRevLett.127.026401,Vano2021} open up fundamentally new perspective to study quantum criticality and emergent phenomena.

In this article, we analyze the heavy fermion physics in a real-space Moir\'{e} superlattice in two fashions: i) by introducing a twist-angle between the itinerant conduction band sites and the correlated $f$-electrons and ii) by systematically depleting the correlated $f$-electron sites, effectively increasing the unit cell of the MSL. By employing a periodic Anderson lattice model (PAM) with periodically diluted $f$ orbitals and solving the problem within dynamical mean field theory (DMFT) \cite{DMFT_1, DMFT_2}, we discover the flat conduction band arising in such a heavy-fermion MSL causes a dramatic impact on the Kondo and coherence scale. Most strikingly,  our study reveals the existence of a third low-energy scale, \( T_\text{Re} \), which lies between \( T^\text{loc}_0 \) and \( T^\text{glob}_0 \). This scale represents the temperature at which Kondo correlations originating from local screening at \( T \approx T^\text{loc}_0 \) induce another magnetic moment that is spatially spread within the unit cell. The underlying mechanism connects with recently discussed reentrant Kondo physics within a specific single impurity model \cite{ReentrantKondo_1, ReentrantKondo_2}, generated through a long wavelength interference scattering through the Moir\'{e} unit cell. 
By systematically varying the spatial separation of the correlated orbitals or the twist angle and analyzing the resulting spectral functions and magnetic properties, we demonstrate the intricate interplay between local and global screening processes. Our findings suggest that the second-stage Kondo temperature \( T^\text{glob}_0 \) can be exponentially suppressed, particularly at high-symmetry points where the Lieb-Mattis theorem predicts a finite ground state spin component. We argue that the existence of \( T_\text{Re} \) is necessary to consistently bridge these predictions with well-established properties of the single impurity limit for well-isolated correlated orbitals. We study the dependence of the low-energy temperature scales on the model parameters in the presence and absence of particle-hole symmetry and numerically extract the characteristic length scale of the Kondo cloud. 

\section{Model and method}
\label{sec:materials_and_methods}

\begin{figure*}[!t]
    \centering
    \includegraphics[width=0.9\linewidth]{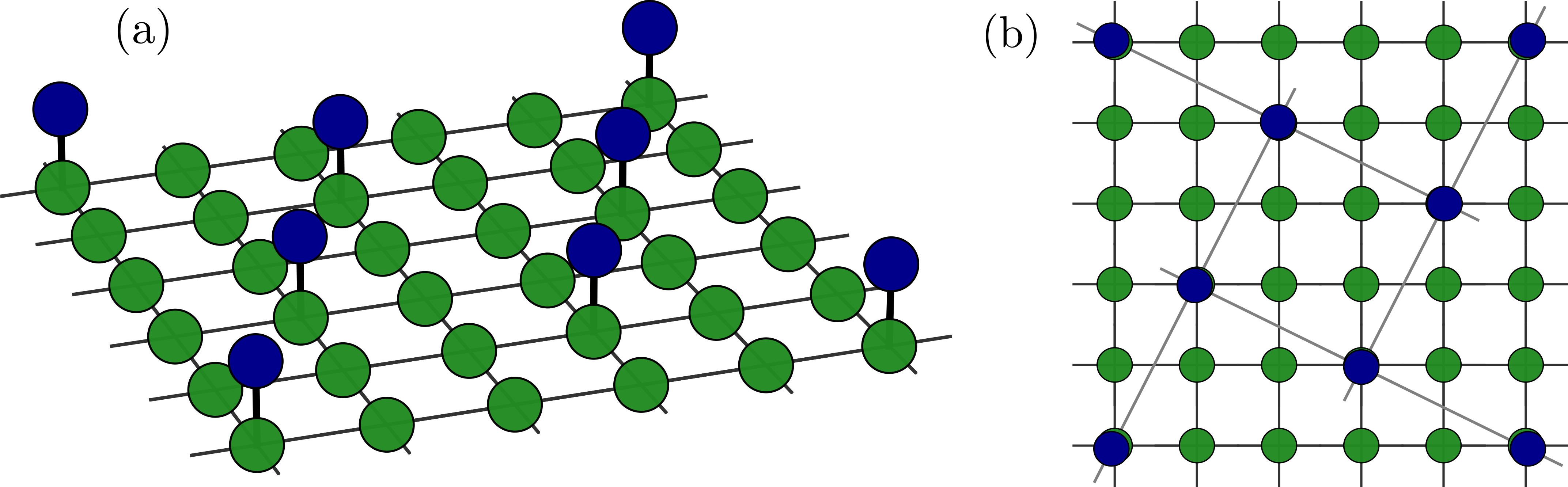}
    \caption{\textbf{Schematic model.}
    (a) side and (b) top view of a specific realization of the model. Green circles represent uncorrelated $c$-orbitals, allowing hopping between nearest neighbors. Within each unit cell, there exists a single correlated $f$-orbital, denoted by the blue circle. While the distance between $c$-orbitals remains fixed at $\Delta R^c=a$, the separation between $f$-orbitals can be greater. Additionally the $f$- and $c$-lattice can be rotated by a twist angle $\phi$.
    } 
    \label{fig:7}
\end{figure*}

We investigate the periodic Anderson model (PAM) on a MSL by diluting periodically the correlated $f$-orbital sites. The model, schematically depicted in Fig.~\ref{fig:7}, is described by the Hamiltonian:
\begin{align}
    \label{eq:H}
    H = H^\text{c} + H^\text{f} + H^\text{hyb}.
\end{align}
The components of this Hamiltonian describe various aspects of the system: the non-interacting bands denoted as $H^\text{c}$, the local interacting $f$-orbitals as $H^\text{f}$, and the hybridization term as $H^\text{hyb}$. The real-space representation of the Hamiltonian is given by:
\begin{align}
    \label{eq:Hc}
    H^\text{c} &= \sum_{i, j, \sigma} [t_{ij} c^\d_{i\sigma}c^{\pd}_{j\sigma}+ \text{h.c.}]\;,\\
    \label{eq:Hf}
    H^\text{f} &= \sum_{l,\sigma} [\epsilon^\text{f} n^\text{f}_{l\sigma} + \frac{U}{2} n^\text{f}_{l\sigma} n^\text{f}_{l\bar{\sigma}}]\;, \\
    \label{eq:Hhyb}
    H^\text{hyb} &= \sum_{l,\sigma}
    [V f^\d_{l\sigma}c^{\pd}_{l\sigma} + \text{h.c.}]\;,
\end{align}
where $n^\text{f}_{l\sigma} = f^\dag_{l\sigma}f^{\pd}_{l\sigma}$.
The creation and annihilation operators $c^\dag_{i\sigma}$ ($c_{i\sigma}$) generate (annihilate) electrons with spin $\sigma$ at the non-interacting $c$-orbital at the real-space lattice site $i$, while $f^\dag_{l\sigma}$ ($f_{l\sigma}$) create (annihilate) an electron with spin $\sigma$ in the correlated $f$-orbital at site $l$.
The tunneling amplitude between $c$-orbitals $i$ and $j$ is denoted as $t_{ij}$, where we define  $\epsilon^c=t_{ii}$. The Coulomb interaction strength of the $f$-orbitals is represented by $U$, and both subsystems hybridize locally with strength $V$.
In our MSL model, the correlated $f$-orbital sites are not present at every lattice site. This effectively enlarges the unit cell of the Hamiltonian, resulting in a greater total number of $c$-orbital sites compared to $f$-orbital sites ($N_c > N_f$). As illustrated in Fig.~\ref{fig:7}(a) and (b), we assume that the lattice structures of the $c$- and $f$-subsystems are identical. However, the spacing between the two types of orbitals may differ, with $\Delta R^c = a \leq \Delta R^f$. Additionally, the unit vectors of the respective lattices might be rotated relative to each other by a twist angle $\phi$.
In this work we focus on a square lattice in two dimensions with hopping elements between nearest neighbors only, and use the band center, $\epsilon^c$, for particle and hole doping respectively.

\subsection{Connection to moir\'e superlattices and emergence of local moments}
\label{sec:moire_connection}

Moir\'e superlattices and van der Waals heterostructures provide a versatile platform in which localized magnetic moments coexist with itinerant
electrons, giving rise to effective Anderson- or Kondo-lattice physics.
Two conceptually distinct realizations have been established experimentally.

First, intrinsic van der Waals heavy-fermion materials host localized $f$-electron moments that hybridize with dispersive conduction bands.
Recent experiments on Ce-based layered compounds have demonstrated two-dimensional heavy-fermion behavior and Kondo coherence in atomically
thin systems \cite{BGJang2022,VAPosey2024}.
In this case, the separation between localized and itinerant degrees of freedom is microscopic, and the periodic Anderson model provides a
direct description.

Second, in Moir\'e heterostructures based on transition-metal dichalcogenides (TMDs), strong Coulomb interactions within narrow Moir\'e minibands
can drive Mott localization in one layer, while itinerant carriers remain in another band or layer.
This leads to an emergent array of localized moments coupled to conduction electrons, realizing a gate-tunable Kondo lattice,
as observed in MoTe$_2$/WSe$_2$ bilayers \cite{Zhao2023,Guerci2023}.
Related correlated states have also been reported in other layered systems such as 1T-TaS$_2$/1H-TaS$_2$ heterostructures \cite{Vano2021}.

A third class of systems is provided by twisted graphene multilayers, where correlated insulating states and strong band renormalization lead to emergent local moments coupled to itinerant bands.
Several works have proposed mappings of magic-angle twisted bilayer
and trilayer graphene to Kondo-lattice-type models, often with enlarged internal symmetries such as SU(4) arising from spin and valley degrees of freedom \cite{PhysRevLett.127.026401,Song2022,Shi2022,Hu2023,Chou2023}.
Experimental signatures consistent with local-moment formation and heavy-fermion phenomenology have been reported in thermodynamic and
spectroscopic probes \cite{Ghosh2025,Merino2025,zhang2025,calugaru2024,merino2024,batlleporro2024}.

The effective lattice model studied in the present work should be viewed in this broader context.
It captures a situation in which localized orbitals form a periodic lattice and hybridize with itinerant electrons whose band structure is modified
by a superlattice potential, leading to Brillouin zone backfolding and multiple conduction channels per momentum.

However, we emphasize that our model is not intended as a microscopic description of twisted graphene systems.
In particular, in Moir\'e graphene the localized moments are emergent degrees of freedom associated with flat bands on a triangular Moir\'e lattice,
while the itinerant electrons reside in a continuum description and the effective models often exhibit approximate SU(4) symmetry.
In contrast, our model assumes a lattice of localized $f$ orbitals hybridizing with a tight-binding conduction band and does not incorporate such enlarged internal symmetries.

Instead, our approach should be regarded as a minimal lattice realization of multi-channel hybridization induced by Moir\'e backfolding.
It is therefore most directly applicable to systems where localized orbitals coexist with lattice conduction electrons,
such as van der Waals heavy-fermion materials and TMD Moir\'e heterostructures, while capturing universal interference effects that
are expected to arise whenever multiple conduction channels couple coherently to a periodic array of local moments.

\subsection{Numerical details}
To tackle this complex model in the translational invariant form (PAM), we employ DMFT \cite{DMFT_1, DMFT_2}. Details on how the involved Greens functions can be computed in an efficient way can be found in Appendix~\ref{sec:Gf_and-DMFT}.
In order to solve the self-consistent single-impurity Anderson impurity model (SIAM)
we employ the Numerical Renormalization Group (NRG) technique,
as implemented in the NRG-Lubljana interface \cite{TRIQS_NRG, TRIQS_NRG2} to the TRIQS open-source package \cite{TRIQS}.
In our NRG calculations, we utilized a $z$-averaging technique over $N_z=4$ values of the twist parameter \cite{NRG_discretization}. This approach was combined with an enhanced discretization scheme \cite{TRIQS_NRG} and the discretization parameter was set to $\lambda = 2$, while 2000 states were kept in each iteration.
For computing spectral functions, we employed a complete basis set of the Wilson chain \cite{NRG_Spectra1, NRG_Spectra2} and made use of the self-energy trick \cite{NRG_selfEnergy}. To broaden the spectral features, we employed the procedure described in Ref.~\cite{NRG_broadening}, with a broadening parameter $\alpha = 0.4$.

\section{Results}
\label{sec:results}

\subsection{Local moment revival - Lieb-Mattis limit}
\label{sec:LiebMattis}

In the finite-size Kondo lattice, an intriguing adaptation of the Lieb-Mattis theorem \cite{Lieb1962}, as originally delineated by Shen \cite{Shen1996}, emerges. This theorem pertains to a scenario where $N_f$ local moments interact locally with a half-filled ensemble of conduction electrons on a bipartite d-dimensional lattice. In this setup, with $N_c\geq N_f$ sites on the lattice, and considering a finite Hubbard-type interaction among the conduction electrons, the theorem elucidates the total $S_z$ component of the ground state:

\begin{align}
S^\text{tot}_z(T=0) = \frac{1}{2}|N_c^\text{A} - N_c^\text{B} + N_f^\text{B} - N_f^\text{A}|,
\label{eq:LiebMattis}
\end{align}
where $N_c^\text{A/B}$ ($N_f^\text{A/B}$) referes to the number of $c$-sites ($f$-sites) on the A or B lattice respectively.
This expression, detailed as theorem VI in Shen's work \cite{Shen1996}, signifies a pivotal insight into the system's magnetic properties.

In the absence of interactions among the conduction electrons, degeneracy of the ground state, apart from the trivial ($2S^\text{tot}_z+1$)-fold degeneracy, is only precluded in the dense configuration, where $N_f=N_c$. However, even if degeneracy is present, one of the ground states invariably aligns with the $S^\text{tot}_z$ sector specified by Eq. \eqref{eq:LiebMattis}.

Additionally, Titvinidze et al. \cite{Depleted_PAM_3} illustrated that the theorem could be applied to the regularly depleted Kondo lattice in one and two dimensions. Using several techniques, including the Density Matrix Renormalization Group (DMRG), DMFT  and perturbation theory, their demonstration revealed that even when the conduction electrons are non-interacting, the ground state remains unique.

For the defined model,
the Lieb-Mattis theorem, and Eq.~\eqref{eq:LiebMattis} respectively apply in the case of particle-hole symmetry, $\epsilon^c=0$, $\epsilon^f=-U/2$. For even $f$-orbital separation, $\Delta R^f=2n\, a$, where $n\in \mathbb{N}$, and $\phi=0$ we find $N^\text{A}_c-N^\text{B}_c=0$ and $N^\text{A}_f-N^\text{B}_f=N_f=N_\text{u}$, with $N_\text{u}$ representing the number of unit cells in the system. Consequently, Eq.~\eqref{eq:LiebMattis} predicts $S^\text{tot}_z=\frac{N_\text{u}}{2}$.
For these highly symmetric model parameters the non-interacting band structure, $U=0$, displays a completely flat band right at the Fermi energy. 

Within paramagnetic single-site DMFT in the thermodynamic limit, as $N_\text{u}\to \infty$, this implies that the effective impurity problem, representing the unit cell of the lattice model, should remain in the local moment fixed point. If ferromagnetic correlations beyond a mean-field analysis are incorporated, these local moments will aggregate into a large moment with $S_z=\frac{N_\text{u}}{2}$.

We leverage the predictions of the Lieb-Mattis theorem for two purposes: firstly, for benchmarking the numerical results, especially when dealing with large unit cells; and secondly, to address a question arising when attempting to reconcile the prediction of Eq.~\eqref{eq:LiebMattis} with the limit of isolated impurities for exceedingly large unit cells, $\Delta R^f\to \infty$. While the Lieb-Mattis theorem predicts a local moment per unit cell, one expects to recover the physics of the SIAM when local moments are well separated. In this scenario, the local impurity spin becomes screened on the characteristic energy scale known as the Kondo temperature $T_\text{K}$, signaling the crossover from local moment to the strong coupling singlet fixed point.

As we demonstrate in the following, this apparent discrepancy is resolved by the emergence of a new "below Kondo" energy scale, $T_\text{Re}\leq T_\text{K}$, at which a spin degree of freedom reemerges within the unit cell of the lattice model.

Typically, the unit of energy in impurity problems is defined by $\Gamma=\pi V^2 \rho^c(0)$, where $\rho^c(0)$ represents the density of states of the conduction band electrons precisely at the Fermi energy.

However, due to the Van Hove singularity in the density of states of a two-dimensional half-filled ($\epsilon^c=0$) square lattice precisely at the Fermi energy, we establish $V=1$ as the reference in this subsection. Additionally, we set the bandwidth $D$ to $D/V=2$, and particle-hole symmetric impurity parameters are achieved by setting $\epsilon^f=-U/2$.

\begin{figure*}[t]
    \centering
    \includegraphics[width=0.99\linewidth]{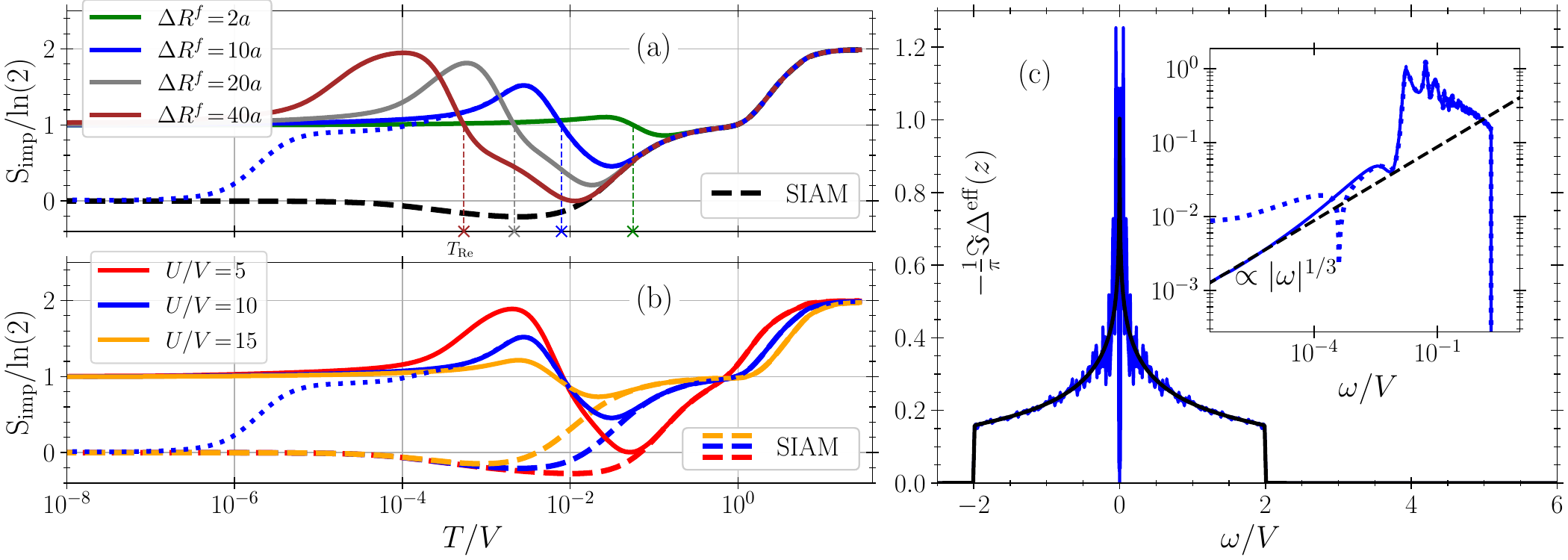}
    \caption{\textbf{Spin revival in the Lieb-Mattis limit.} DMFT results for PH-symmetric model parameters, $D/V=2$, $\epsilon^c=0$, $\epsilon^f=-U/2$, and even $f$-orbital separations $\Delta R^f=2n a$ with $\phi=0$.
    Panels (a) and (b) depict the impurity-induced entropy $\text{S}_\text{imp}$ obtained for the effective SIAM after reaching self-consistency. In panel (a), a fixed value of $U/V=10$ is utilized, with solid lines representing different $f$-orbital separations. Vertical dashed lines indicate the definition of the energy scale $T_{\mathrm{Re}}$. Conversely, in panel (b), the solid lines indicate various strengths of interactions, while $\Delta R^f=10a$ remains fixed. Dashed lines in both panels correspond to the respective SIAM, while the dotted line corresponds to a slight deviation from the PH-symmetric point for $\Delta R^f=10a$ and $U/V=10$, with $\epsilon^c/D=2\cdot 10^{-4}$. Panel (c) displays the imaginary part of the effective DMFT medium for $\Delta R^f=10a$ and $U/V=10$ (solid blue line) in comparison with the bare $c$-orbital DOS $V^2\rho^c(\omega)$ (solid black line). The same spectra is plotted on a logarithmic scale for $\omega>0$ in the inset in comparison with the same spectra calculated for slight deviation from the PH-symmetric point with $\epsilon^c/D=2\cdot 10^{-4}$ (dotted blue line). The dashed line in the inset of panel (c) is proportional to $|\omega|^{1/3}$.
    }
    \label{fig:1}
\end{figure*}

In Fig.~\ref{fig:1}, we illustrate the impurity-induced entropy $S_\text{imp}$. As customary in the NRG, it is defined as the difference between the total system, which includes the impurity coupled to the Wilson chain, and the reference system, i.e., the Wilson chain without the impurity. While the solid lines represent the DMFT solution of the lattice model, the dashed lines depict the results for the corresponding SIAM on a two-dimensional square lattice.

In Fig.~\ref{fig:1}(a), the $f$-orbital separation, $\Delta R^f/a$, is progressively increased, while keeping the local impurity parameter fixed at $U/V=10$. Consequently, only a single SIAM representation is present.
Contrary to that, in Fig.~\ref{fig:1}(b) the strength of interaction $U/V$ was varied across, while keeping $\Delta R^f=10a$ fixed. As the parameters of the $f$-orbital are not constant in this case the respective SIAM representation differs as well, indicated by the dashed line with the corresponding color.

In the high-temperature regime, where $T/V\gg 1$, the impurity-induced entropy per unit cell of the DMFT solution and the SIAM correspond to the free orbital fixed point (FP), exhibiting $S_\text{imp}=2\ln(2)$.
Upon reducing the temperature, a transition occurs to the local moment FP, characterized by $S_\text{imp}=\ln(2)$, occurring on a scale of $T\approx U$. Subsequently, Kondo screening initiates around $T\approx T_{\text{K}}\propto\exp{(-U/V)}$.

In the case of SIAM representations (dashed lines), the impurity-induced entropy steadily decreases, reaching a minimum with $S_\text{imp}<0$, before attaining the stable strong coupling FP with $S_\text{imp}=0$ as $T\ll T_\text{K}$. The occurrence of negative $S_\text{imp}$ values, elucidated in \cite{NegSimp}, is well-understood and arises whenever sharp peaks in the DOS of the medium exist near the Fermi energy, such as the Van Hove Singularity in this instance.

Conversely, the DMFT solution begins to deviate from the SIAM on a scale $T_\text{Re}$, marked by an increase in $S_\text{imp}$ upon further temperature reduction. Fig.~\ref{fig:1}(a) depicts that $T_\text{Re}$ diminishes with increasing $f$-orbital separation $\Delta R^f$, while panel (b) illustrates that $T_\text{Re}$ remains independent of the interaction strength $U/V$. For $T<T_\text{Re}$, the impurity-induced entropy ascends until it surpasses $S_\text{imp}\geq \ln(2)$, after which it descends again towards the stable local moment FP as $T\to 0$, as predicted by the Lieb-Mattis theorem. The greater the $f$-orbital separation and the weaker the interaction strength, the closer $S_\text{imp}(T<T_\text{Re})$ approaches the maximum value of $2\ln(2)$, corresponding to the free orbital FP.

Figure~\ref{fig:1}(c) exhibits the imaginary component of the effective medium for $\Delta R^f=10a$ and $U/V=10$, corresponding to the solid blue lines in Fig.~\ref{fig:2}(a) and (b). At higher energies, the medium closely resembles the DOS of the square lattice (solid black line), embellished with minor oscillations. However, instead of the Van Hove singularity, $\Im \Delta^\text{eff}(z)$ showcases a sharp dip precisely at $\omega=0$. As discernible from the inset, where the spectrum is displayed on a double logarithmic scale, the DOS of the effective medium exhibits a power-law behavior $\propto |\omega|^{1/3}$ at lower frequencies, which does not depend on the specific choice of $\Delta R^f$ or $U/V$ \cite{MOPAM}.

The physics of the SIAM in the presence of a pseudo-gap DOS, characterized by an exponent $r$, $\rho^c(\omega)\propto |\omega|^r$, has been extensively studied \cite{SIAM_Gap1, SIAM_Gap2, SIAM_Gap3, SIAM_Gap4, SIAM_Gap5, SIAM_Gap6, SIAM_Gap7, SIAM_Gap8}. In cases of PH-symmetry, the local moment FP remains stable for $r\geq 0.5$, while a critical Kondo coupling strength $J_c(r)$ has been identified for $r<1/2$ \cite{SIAM_Gap4, SIAM_Gap7}, governing the transition from a stable local moment FP to a stable strong coupling FP above $J_c(r)$. The critical point at $J=J_c(r)$ is characterised by an interacting non-Fermi-liquid FP with substantial local moment fluctuations \cite{SIAM_Gap7, SIAM_Gap6, SIAM_Gap8} and notable $\omega/T$ scaling behavior \cite{SIAM_Gap6}, consistent with experimental observations in various heavy Fermion compounds at criticality \cite{LQCP_1, LQCP_2, LQCP_3}.

However, despite the pseudo-gap of the effective medium of the DMFT solution is described by an exponent $r^\text{f}=0.33<1/2$, the effective coupling generated through the DMFT self-consistency cycle consistently remains below the critical coupling strength, ensuring the stability of the local moment fixed point as predicted by the Lieb-Mattis theorem. For a more detailed discussion on this we refer the reader to Sec.~4.3.2 of Ref.~\cite{MIAM}.

While initial studies of the pseudo-gap SIAM generally assumed the power-law dependency to span the entire bandwidth of the conduction band DOS, within the DMFT solution of the lattice model under consideration here, such behavior is observed only at low frequencies, $\omega\approx T_\text{Re}$, which can be smaller than $T_\text{K}$ of the corresponding SIAM.

In this specific scenario, $T_\text{Re}<T_\text{K}$, the physics emerging from the DMFT self-consistency cycle in the lattice model connects with a phenomenon termed "reentrant Kondo physics," recently discussed in great detail within the single impurity context \cite{ReentrantKondo_1, ReentrantKondo_2}. When a quantum impurity interacts with a metallic host featuring a sharp gap of width $\Delta_\text{gap}<T_\text{K}$ around the Fermi energy, the typical Kondo quenching process at $T\approx T_\text{K}$ is succeeded by a secondary sequence of SIAM fixed points, namely, free orbital FP $\to$ local moment FP $\to$ strong coupling FP. The emergence of the second-stage free orbital FP occurs at approximately $T\approx\Delta_\text{gap}<T_\text{K}$, while the second-stage Kondo temperature decreases exponentially with the depth of the gap \cite{ReentrantKondo_1, ReentrantKondo_2}.

For the lattice model in the Lieb-Mattis limit, the DMFT solution displays a complete pseudo-gap, vanishing precisely at $\omega=0$, thereby stabilizing the local moment fixed point. However, slight deviations from the particle-hole symmetric point render the Lieb-Mattis theorem inapplicable, potentially allowing for a second-stage Kondo screening. Fig.\ref{fig:1}(a-c) present the DMFT solution for parameters $\Delta R^f=10a$, $U/V=10$, and a slightly shifted $c$-band center $\epsilon^c/D=2\cdot 10^{-4}$ as dotted blue lines. Indeed, the inset of Fig.\ref{fig:1}(c) reveals the presence of a very sharp gap-like feature around the Fermi energy; however, it now has a finite depth, leading to a second-stage Kondo quenching and a stable strong coupling fixed point as revealed in panels (a-b).

The Kondo breakdown, manifesting as the vanishing of \(T^\text{glob}_0\), is a direct consequence of perfect mirror symmetry in the conduction band structure after it is back-folded into the reduced Brillouin zone. 
When the \(f\)-orbitals hybridize with the conduction band~\cite{MOPAM}, 
this symmetry leads to a completely flat band, indicating that interference in the hybridization is destructive. 
For more details on the physical mechanism behind this Kondo breakdown scenario, 
we refer to Sec.~\ref{sec:screening_scales}, Appendix~\ref{sec:mechanism} and reference Ref.~\cite{MOPAM}.

\subsection{Reemergence energy scale and local screening cloud}

In this subsection, we discuss the dependency of the crossover energy scale $T_\text{Re}$ on the model parameters.

As $T_\text{Re}$ represents the temperature at which $S_\text{imp}(T)$ increases upon lowering $T$, we extract $T_\text{Re}$ from $S_\text{imp}$ at the temperature where it crosses $\ln(2)$ from below: $S_\text{imp}(T_\text{Re}) = \ln(2)$, with $S_\text{imp}(T \gtrsim T_\text{Re}) < \ln(2)$.

\begin{figure}[t]
    \centering
    \includegraphics[width=0.99\linewidth]{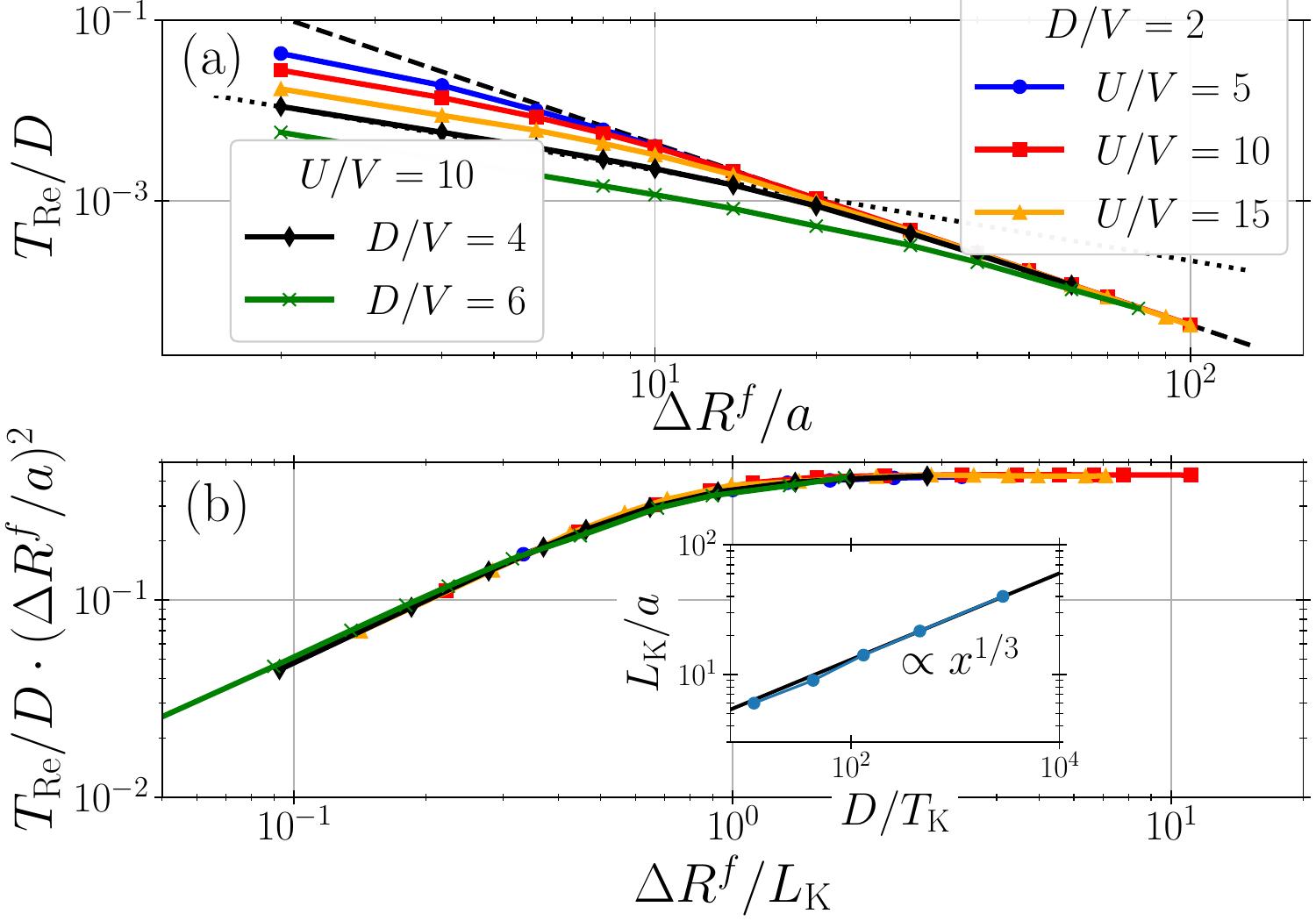}
    \caption{\textbf{Reemergence energy scale and Kondo screening cloud.} (a) Low-energy temperature scale $T_\text{Re}/D$ as a function of the $f$-orbital separation $\Delta R^f = 2n a$, $\epsilon^c = 0$, and $\epsilon^f = -U/2$. Various colors and markers denote different combinations of $U/V$ and $D/V$.
    The dotted black line represents the dependency $\propto (\Delta R^f)^{-1}$, while the dashed black line is proportional to $(\Delta R^f)^{-2}$.
    (b) Same as panel (a) but rescaled by $(\Delta R^f/a)^2$ and plotted as function of $\Delta R^f/L_\text{K}$, where $L_\text{K}$ denotes the $f$-orbital separation at which $T_\text{Re}$ crosses over from $(\Delta R^f)^{-1}$ to $(\Delta R^f)^{-2}$ dependency. The Inset of panel (b) depicts the dependency of $L_\text{K}$ on the corresponding SIAM Kondo temperature $T_\text{K}$ on a double logarithmic scale where the black solid line corresponds to $\propto x^{1/3}$.}
    \label{fig:3}
\end{figure}

Figure~\ref{fig:3}(a) illustrates the dependence of $T_\text{Re}$ on $\Delta R^f$, with twist angle $\phi=0$, and various combinations of $U/V$ and $D/V$. Particle-hole (PH) symmetry is realized by setting $\epsilon^c = 0$ and $\epsilon^f = -U/2$.

For small $f$-orbital separations $\Delta R^f$, the reemergence scale $T_\text{Re}$ follows a $(\Delta R^f)^{-1}$ dependency (dotted black line); however, the corresponding prefactor strongly depends on the model parameters.

In contrast, in the limit of large $f$-orbital separations $\Delta R^f \to \infty$, the temperature scale $T_\text{Re}$ depends only on the bandwidth $D$, such that $T_\text{Re}/D$ becomes a universal function and follows a $(\Delta R^f)^{-2}$ dependency (dashed black line). Importantly, $T_\text{Re}$ is independent of the local $f$-orbital parameters $U$ and $V$ in this limit.

In the context of single impurity models, the Kondo screening is often characterized by a specific length scale known as the Kondo screening cloud \cite{KondoCloud_1}. This cloud represents the spatial region around the magnetic impurity where conduction electrons become correlated to effectively screen the impurity's magnetic moment, and has been observed in experiment recently \cite{KondoCloud_2}.

The results of the lattice model presented in Fig.~\ref{fig:3}(a) can be interpreted using this picture of a Kondo screening cloud length scale within the corresponding SIAM limit.
For this we define the scale $L_\text{K}$ as the $f$-orbital separation at which $T_\text{Re}$ crosses over from $(\Delta R^f)^{-1}$ to $(\Delta R^f)^{-2}$ dependency.

For $\Delta R^f \ll L_\text{K}$, the Kondo screening cloud is much larger than the unit cell, causing the screening clouds of different $f$-orbitals to overlap. The magnitude of this overlap and the size of the respective screening clouds depend on the Kondo temperature and the local $f$-orbital parameters.

In the opposite limit, $\Delta R^f \gg L_\text{K}$, the screening clouds fit into the unit cell, enabling independent Kondo screening of the respective $f$-orbitals. Hence, from a local perspective of the $f$-orbitals, $L_\text{K}$ represents the crossover from lattice to impurity behaviour. In this limit, the overlap of different screening clouds does obviously not depend on local impurity parameters, leaving the bandwidth as the only relevant energy scale.

Figure~\ref{fig:3}(b) depicts the same data as shown in panel (a) but rescaled by $(\Delta R^f/a)^2$ and plotted as a function of $\Delta R^f/L_\text{K}$. All data points collapse onto a single universal curve, confirming the hypothesis that $L_\text{K}$ is the only relevant length scale of the system and characterizes the crossover from collective to independent Kondo screening of the local $f$-orbital magnetic moments. However, it is important to note that this does not imply a crossover from lattice to impurity physics for the whole system. While the Kondo-screened strong coupling singlet fixed point is stable in the corresponding SIAM, these Kondo correlations lead to the reemergence of the magnetic moment in the lattice model on the scale denoted by $T_\text{Re}$.

The inset of panel (b) shows the dependency of $L_\text{K}$ on the Kondo temperature $T_\text{K}$ of the corresponding SIAM on a double logarithmic scale. The black solid line in the inset indicates a power-law relation $L_\text{K} \propto (D/T_\text{K})^{r}$ with $r_\text{DMFT} = \frac{1}{3}$. Interestingly, this exponent deviates from the originally predicted value of $r_\text{theory} = 1$ \cite{KondoCloud_1}.

The physical picture emerging so far is as follows: In the lattice model, there are three relevant energy scales, which we denote as $T^\text{loc}_0$, $T_\text{Re}$, and $T^\text{glob}_0$.
Considering the results presented in Fig.~\ref{fig:1}, $T^\text{loc}_0$ denotes the first-stage Kondo screening of the local moment at the impurity site, while $T^\text{glob}_0$ refers to the second-stage Kondo quenching of the spin degree of freedom that emerges at $T_\text{Re}$.
If the Lieb-Mattis limit is applicable, the second-stage Kondo screening disappears ($T^\text{glob}_0=0$), and the second-stage local moment FP is stable.
For large $\Delta R^f$, we have $T_\text{Re}\ll T^\text{loc}_0$, and Fig.~\ref{fig:3}(a) suggests $T^0_\text{Re}\to 0$ for $\Delta R^f\to \infty$.
Consequently, we can identify $T^\text{loc}_0$ with the SIAM Kondo temperature in the limit of large $f$-orbital separation, which is the only relevant energy scale for $\Delta R^f\to\infty$.
For finite but large $\Delta R^f$, an effective SIAM description is only valid for intermediate temperatures ($T>T_\text{Re}$).

In order to further quantify this scenario, we will now discuss the thermodynamics of the effective magnetic moment present within each unit cell of the lattice model, but distinguish between the local moment $\mu^2_\text{loc}$ right at the impurity site and the global magnetic moment $\mu^2_\text{glob}$ corresponding to the whole unit cell. For details on the numerical calculation of $\mu^2_\text{loc}$ and $\mu^2_\text{glob}$ we refer to Appendix~\ref{sec:method_loc_vs_glob}.

\begin{figure}[t]
    \centering
    \includegraphics[width=0.99\linewidth]{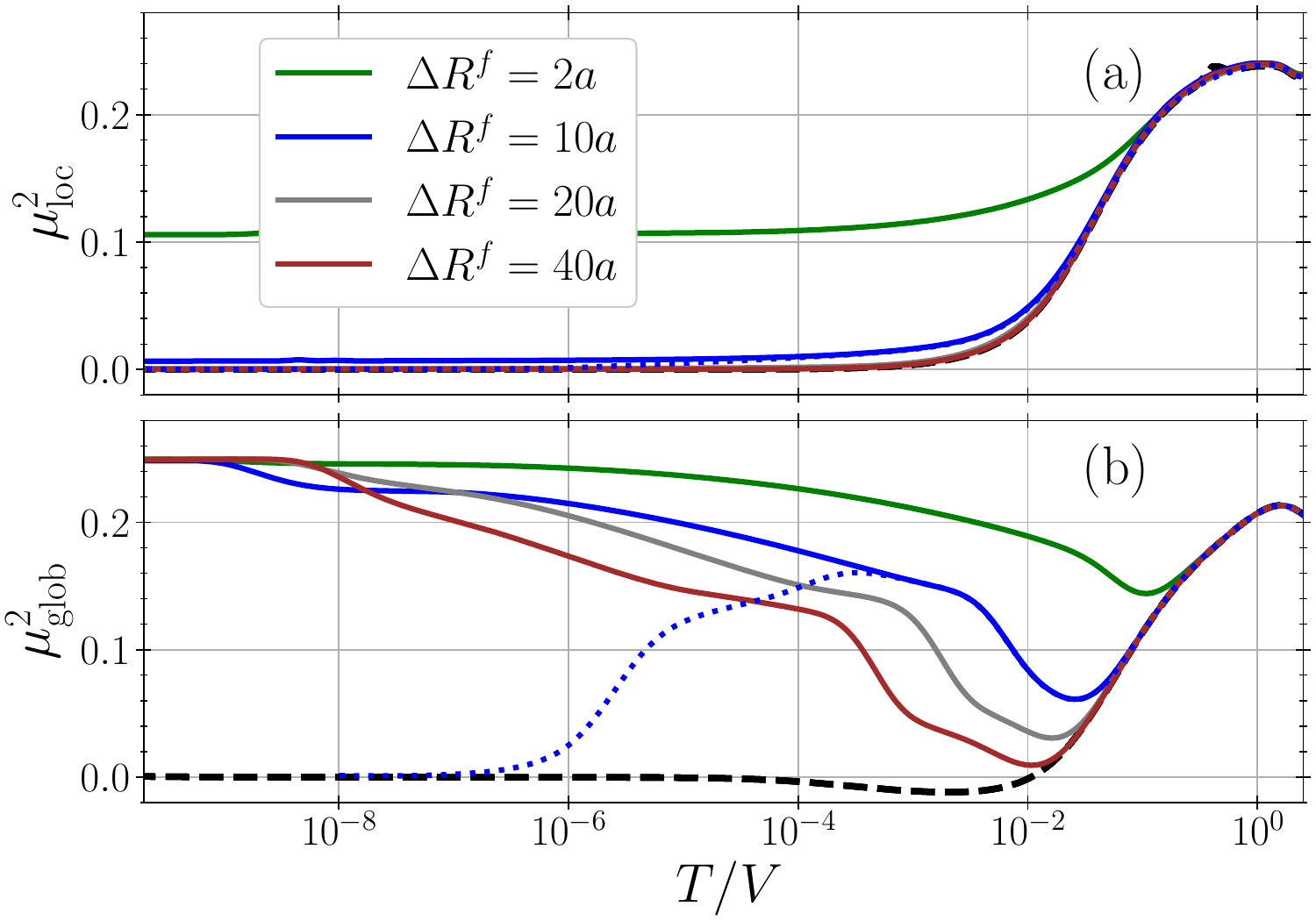}
    \caption{\textbf{Temperature dependence of local and global magnetic moment.} Effective magnetic moment of (a) the local impurity site and (b) the total system for the identical parameter set as in Fig.~\ref{fig:2}(a). The dashed line represents the respective SIAM, while the dotted line represents a slight deviation from the PH-symmetric point for $\Delta R^f=10a$, with $\epsilon^c/D=2\cdot 10^{-4}$.
    }
    \label{fig:2}
\end{figure}

Figure~\ref{fig:2} depicts the local and global effective magnetic moment $\mu^2_\text{loc}$ and $\mu^2_\text{glob}$ as functions of temperature for the same set of parameters as Fig.~\ref{fig:1}(a).
At high temperatures, we have $\mu^2_\text{loc} \approx \mu^2_\text{glob}\approx 0.25$, corresponding to the free spin-$\frac{1}{2}$ moment right at the local $f$-site.
Around $T\approx T^\text{loc}_0$, both the local and global effective moments start to decrease, signaling the onset of local Kondo screening, consistent with the corresponding SIAM result (black dashed lines).
The local magnetic moment $\mu^2_\text{loc}$ continues to decrease and saturates at a constant value in the zero temperature limit, $\mu^2_\text{loc}(T\to 0)>0$. The larger the $f$-orbital separation, $\Delta R^f$, the later $\mu^2_\text{loc}$ starts to saturate, resulting in a smaller value for $T\to 0$. For $\Delta R^f=20a$, the DMFT result for $\mu^2_\text{loc}$ closely resembles the corresponding SIAM result (black dashed line).

The behavior of $\mu^2_\text{glob}$ differs significantly. Similar to the increase of $S_\text{imp}$ in Fig.~\ref{fig:1} right after the onset of the first-stage Kondo screening, $\mu^2_\text{glob}$ also starts to increase at $T\approx T_\text{Re}$ and approaches $\mu^2_\text{glob}=0.25$ for $T\to 0$, consistent with the Lieb-Mattis theorem. For larger $\Delta R^f$, the deviation from the corresponding SIAM result appears at smaller temperatures, corresponding to a smaller $T_\text{Re}$.

We also include the result for the slightly shifted $c$-band center, $\epsilon^c/D=2\cdot 10^{-4}$ and $\Delta R^f=10a$, depicted as a dotted blue line in Fig.~\ref{fig:3} (a) and (b). While the difference between the PH-symmetric (solid blue) and PH-asymmetric (dotted blue) result is barely visible in $\mu^2_\text{loc}$, there is a significant difference in $\mu^2_\text{glob}$ at low temperatures $T<T^\text{glob}_0$, where the second-stage Kondo screening sets in.

\subsection{PH asymmetry and flat conduction bands}
\label{sec:screening_scales}

In this subsection, we concentrate on analyzing the dependency of the low energy screening scales $T^\text{loc}_0$ and $T^\text{glob}_0$ for general set of model parameters, away from the PH symmetric point, and discuss consequences for spectral functions. We consider an asymmetric conduction band, $\epsilon^c\not = 0$, such that the Van Hove singularity is shifted from the Fermi energy at $\epsilon_\text{Fermi}=0$.

In the following, we use $\Gamma=\pi V^2 \rho^c(0)=1$ as our unit of energy, such that the Kondo temperature, $T_\text{K}$, of the corresponding SIAM is nearly constant with respect to variations in the conduction band filling, $\epsilon^c$, and given by $T_\text{K}\propto \exp[-\pi U/(8\Gamma)]$ \cite{TK}.

As the screening scales are related to crossover energy scales, there is no precise definition. However, similar to Wilson \cite{Wilson}, we define these screening scales at the temperature where the effective local and global magnetic moment have reached the reduced value of $\mu^2_\text{loc}(T^\text{loc}_0)=\mu^2_\text{glob}(T^\text{glob}_0)=0.05.$
For comparison, we also extracted the Kondo temperature of the corresponding SIAM in the same way,
$\mu^2_\text{loc}(T^\text{loc}_\text{K})=\mu^2_\text{glob}(T^\text{glob}_\text{K})=0.05$. Note, however, that there is only one relevant energy scale in this case, such that $T^\text{loc}_\text{K}\propto T^\text{glob}_\text{K}$ holds. 

\begin{figure*}[t]
    \centering
    \includegraphics[width=0.99\linewidth]{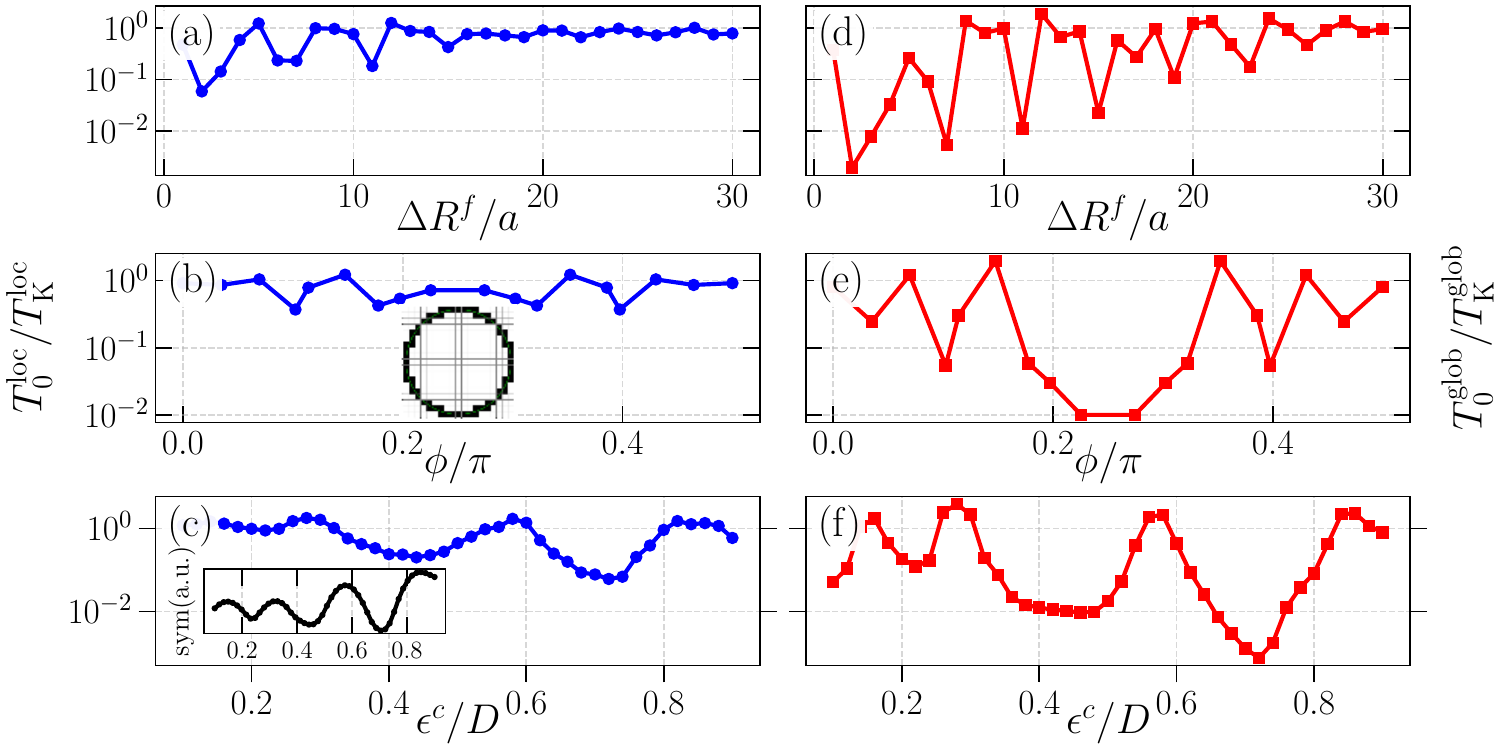}
    \caption{\textbf{Tuning flat band enhancement of correlations.} Dependence of the low-energy screening scales on different control parameters. The left column [(a), (b), and (c)] illustrates the local scale, while the right column [(d), (e), and (f)] displays the global scale.
    In (a) and (d), the \( f \)-orbital separation \(\Delta R^f\) is progressively increased 
    while \(\phi = 0\) and \(\epsilon^c/D = 0.4\).
    In Panels (b) and (e) the twist angle \(\phi\) is varied
    for \(\epsilon^c/D = 0.4\) and \(\Delta R^f/a = 9 \pm \delta\). The Inset in (b) visualizes how $\delta$ is adjusted in order to approximate the circle.
    In panels (c) and (f), we set \(\Delta R^f/a = 3\) with \(\phi = 0\) and continuously increase 
    the band center \(\epsilon^c / D\). Other model parameters are set to $D/\Gamma=5$ and $-2\epsilon^f=U=15\Gamma$.
    }
    \label{fig:5}
\end{figure*}

Figure~\ref{fig:5} illustrates the variation of the ratios 
$T^\text{loc}_0/T^\text{loc}_\text{K}$ [(a), (b), (c)] and $T^\text{glob}_0/T^\text{glob}_\text{K}$ [(d), (e), (f)]
with respect to changes in the model settings, 
while keeping \( U = -2\epsilon^f = 15\Gamma \) and \( D/\Gamma = 5 \) fixed.
In panels (a) and (d), the \( f \)-orbital separation \(\Delta R^f\) is progressively increased 
while \(\phi = 0\) and \(\epsilon^c/D = 0.4\). 
Panels (b) and (e) show how these ratios change upon varying the twist angle \(\phi\) 
for \(\epsilon^c/D = 0.4\) and \(\Delta R^f/a = 9 \pm \delta\), 
where \(\delta\) is chosen so that the discrete set of \( f \)-lattice unit vectors lies approximately 
on a circle of radius \(9a\). 
The inset of Fig.~\ref{fig:5}(b) illustrates these unit vectors on a bitmap, where black pixels correspond 
to the selected unit vectors. 
In panels (c) and (f), we set \(\Delta R^f/a = 3\) with \(\phi = 0\) and continuously increase 
the band center \(\epsilon^c / D\).

Across all cases, the general structure of \(T^\text{loc}_0\) and \(T^\text{glob}_0\) remains similar, 
exhibiting pronounced oscillations as a function of the respective control parameter. 
Notably, these oscillations are much more prominent in \(T^\text{glob}_0\), 
spanning over two orders of magnitude.

The physics emerging here can be understood as a consequence of destructive interference effects 
that arise from an approximate mirror symmetry in the back-folded conduction band \cite{MOPAM}, similar to the Lieb Mattis limit from Sec.~\ref{sec:LiebMattis} where we find full mirror symmetry.
We will illustrate this by focusing on how shifting the band center \(\epsilon^c / D\), 
as shown in panels (c) and (f), impacts the results:

For the standard PAM (\(\Delta R^f = 1a\) and \(\phi=0\) - not shown), the dependence of \(T^\text{loc}_0\) 
and \(T^\text{glob}_0\) on \(\epsilon^c\) is rather featureless~\cite{PAM_1, PAM_2}. 
By contrast, panels (c) and (f) reveal strong oscillations when \(\Delta R^f > 1a\). 
Interestingly, the number of oscillations increases with increasing unit cell size as demonstrated in Appendix~\ref{sec:var_mu}. 
As illustrated in Fig.~\ref{fig:1}, the original (\(V=0\)) band structure of the 
\(c\)-orbitals must be folded back into the Brillouin zone \(\text{Bz}^f\) of the \(f\)-subsystem 
for \(\Delta R^f > 1a\). 
Consequently, the \(f\)-orbitals effectively hybridize with multiple bands, which can intersect 
or touch each other. 
By tuning \(\epsilon^c\) so that such crossing points lie near the Fermi energy - leading to 
approximate mirror symmetry around \(\epsilon_\text{Fermi}\) - interference effects become destructive 
and suppress the overall hybridization strength, causing flat bands to appear in the band structure.
With larger \(f\)-orbital separations \(\Delta R^f\), the increased number of bands and crossing points enhances the number of oscillations.

To quantify this idea, we measure the degree of mirror symmetry in a small window around the Fermi energy 
in the original (\(V=0\)) band structure, folded into the Brillouin zone of the \(f\)-subsystem. 
The result is shown as an inset in Fig.~\ref{fig:5}(c). 
Lower values of this measure correspond to a higher degree of mirror symmetry, 
and the qualitative agreement with the oscillations in the low-energy scales is evident. 
For a detailed description of how this mirror symmetry measure is obtained, 
we refer to Appendix~\ref{sec:var_mu}.

By comparing Fig.~\ref{fig:5} (a) and (d) it seems that SIAM limit is essentially reached at intermediate \(f\)-orbital separations \(\Delta R^f\) in terms of local Kondo-screening (\(T^\text{loc}_0/T^\text{loc}_\text{K}\approx 1\)). Nevertheless, the global behavior of the lattice model can still deviate significantly from the fully dilute limit (\(T^\text{glob}_0 / T^\text{glob}_\text{K}\ll 1\)). This discrepancy highlights the crucial role of correlation effects induced in the nominally non-interacting conduction band, which stem from the Kondo-screening of the local moment at the \(f\)-orbital site:
Whenever the global low-energy screening scale is well-separated from the local one ($T^\text{glob}_0\ll T^\text{loc}_0$), the global (second-stage) screening pertains to the Kondo quenching of a spin degree of freedom emerging at temperature $T_\text{Re}$, reflecting the existence of flat bands in the conduction band.

In order to make connection with possible experiments, we present results for the local density of states (DOS) of the $c$-orbitals, and illustrate that $T^\text{glob}_0\ll T^\text{loc}_0$ indeed manifests itself in the appearance of a sharp Kondo resonance in the spectra of the non-interacting orbitals.

\begin{figure*}[t]
    \centering
    \includegraphics[width=0.99\linewidth]{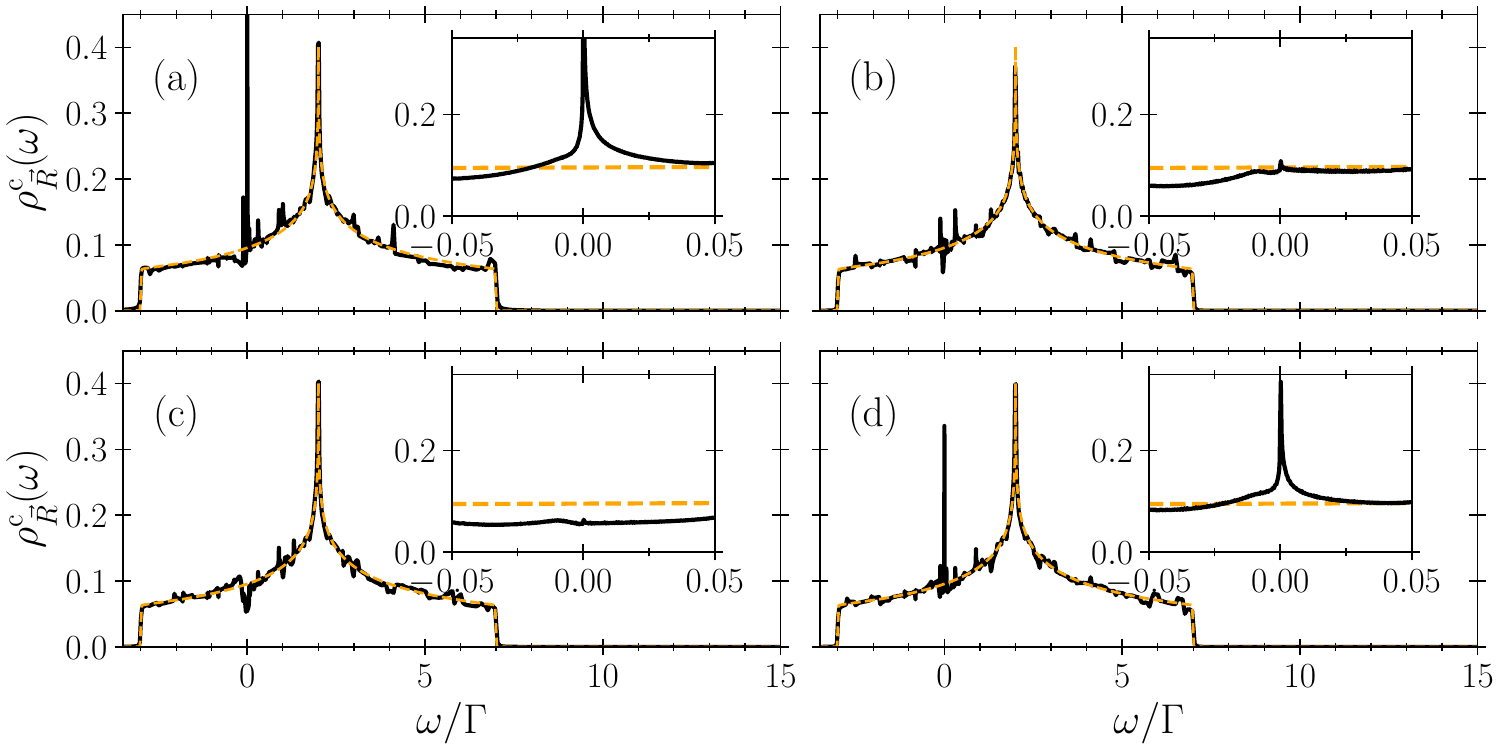}
    \caption{\textbf{Flat-band / Kondo-resonance in the conduction band DOS.} Local DOS of selected $c$-orbitals within the unit cell for a fixed $f$-orbital separation of $\Delta R^f=7a$ with $\phi=0$ and parameters $D/\Gamma=5$, $\epsilon^c/D=0.4$, $U/\Gamma=10$, and $\epsilon^f=-U/2$ (black solid line) in comparison with the original ($V=0$) DOS (orange dashed line). Each panel corresponds to a different $c$-orbital location with respect to the correlated $f$-site: $\vec{R}=(1,0)$ in (a), $\vec{R}=(3,0)$ in (b), $\vec{R}=(2,1)$ in (c), and $\vec{R}=(3,2)$ in (d). Insets provide a zoomed-in view of the spectra around the Fermi energy.
    }
    \label{fig:6}
\end{figure*}

We set the model parameters to $\Delta R^f=7a$, $\phi=0$, $U/\Gamma=10$, $\epsilon^f=-U/2$, $D/\Gamma=5$, and $\epsilon^c/D=0.4$, then employed Eq.~\eqref{eq:Gc_loc} of Appendix~\ref{sec:Gf_and-DMFT} to calculate the local $c$-orbital Greens functions after achieving self-consistency within the DMFT cycle.

Figure~\ref{fig:6} illustrates the local density of states (DOS) of selected $c$-orbitals within the unit cell (solid black lines), compared with the original ($V=0$) DOS (dashed orange lines). Each panel corresponds to different sites at locations $\vec{R}$ relative to the site of the $f$-orbital at $\vec{R}^f=\vec{0}$: $\vec{R}=(1,0)$ in (a), $\vec{R}=(3,0)$ in (b), $\vec{R}=(2,1)$ in (c), and $\vec{R}=(3,2)$ in (d).

At high frequencies, $|\omega|/\Gamma\gg 0$, the DOS of the PAM (black solid lines) at different sites (different panels) are quite similar and closely related to the original ($V=0$) DOS (orange dashed lines) enriched by some oscillations. However, while the DOS at the sites $\vec{R}=(1,0)$ in panel (a) and $\vec{R}=(3,2)$ in panel (b) display a sharp resonance right at the Fermi energy, $\omega\approx 0$, the DOS at sites $\vec{R}=(3,0)$ in panel (b) and $\vec{R}=(2,1)$ in panel (c) are rather featureless. The width of the resonance corresponds to the crossover temperature $T^\text{glob}_0/\Gamma = \mathcal{O}(10^{-4})$ and, consequently, signals the Kondo screening of the reemerged magnetic moment. Interestingly, Fig.~\ref{fig:6} indicates that this spin-$\frac{1}{2}$ degree of freedom is not evenly distributed within the unit cell. While the Kondo resonance is present for sites in the direct neighborhood of the $f$-orbital ($\vec{R}=(1,0)$ in panel (a)) and far away from it ($\vec{R}=(3,2)$ in panel (d)), it is absent for intermediate distances such as $\vec{R}=(3,0)$ and $\vec{R}=(2,1)$ in panels (b) and (c).

The presence of a sharp Kondo resonance in the conduction band validates our interpretation that $T^\text{glob}_0\ll T^\text{loc}_0$ results from Kondo-screening of a magnetic moment distributed within the whole unit cell. However, a detailed analysis of the distribution of this spin degree of freedom and the corresponding quasi-particle interference \cite{MultipleImpOnSurfaces} is left for future work.

\section{Discussion}
\label{sec:discussion}

In this work, we have demonstrated that periodically diluting the correlated \(f\)-electrons in a two-dimensional periodic Anderson model (PAM) leads to a rich Kondo physics with multiple low-energy scales
and Kondo breakdown mechanism arising from destructive interference in the hybridization, accompanied by the occurrence of completely flat bands.
By systematically increasing the distance between neighboring \(f\)-orbitals or rotating the \(f\)- and \(c\)-lattices relative to each other by a twist angle
(thereby realizing a moiré-like superlattices of heavy-fermion layers), we retain an exact description of their multi-band structure and solve the model via DMFT combined with NRG.

One of the main results is the emergence of an additional energy scale \(T_\text{Re}\) between the familiar local Kondo-screening scale \(T^\text{loc}_0\) and the usual lattice-coherence scale \(T^\text{glob}_0\).
Below \(T^\text{loc}_0\), the local moment at the correlated \(f\)-orbital becomes partially screened, inducing correlations in the conduction-band electrons.
Flat bands arising from the moiré-like superlattice further enhance these correlations, causing a second magnetic moment - now spatially extended across the (possibly large) unit cell - to emerge at \(T \approx T_\text{Re}\).
In generic (i.e.\ non-symmetric) situations, this second moment is eventually Kondo-screened at a much lower temperature \(T^\text{glob}_0\).
However, at high-symmetry configurations and fillings, long-wavelength interference scattering across the moiré unit cell drives a true Kondo breakdown.
This destructive interference appears as a vanishing \(T^\text{glob}_0\), consistent with Lieb-Mattis-type arguments that permit a stable local moment in the ground state.
Small deviations from perfect symmetry restore a finite \(T^\text{glob}_0\), placing these observations in the broader context of reentrant Kondo physics with exponentially low screening temperatures.

Even for generic model parameters, we find configurations where \(T^\text{glob}\) is drastically smaller than \(T^\text{loc}\), indicating the existence of flat bands.
Under these circumstances, the spectral functions of some conduction-band orbitals develop a conspicuously narrow Kondo resonance with a characteristic width on the order of \(T^\text{glob}\).
This feature directly reflects the screening of the second, spatially extended local moment that permeates the large moiré supercell.

Experimentally, such a suppressed Kondo-screening scale may manifest in low-temperature probes such as magnetic susceptibility, heat capacity, or scanning tunneling spectroscopy (STS),
where the Kondo resonance of initially uncorrelated sites can be measured directly. These phenomena can be accessed in twisted or stretched van der Waals heterostructures containing \(f\)-electrons,
or in artificially engineered Kondo superlattices, by tuning the real-space periodicities and the relative twist angle between the \(f\)- and \(c\)-lattices.
In so doing, one can promote destructive interference and might drive the system toward a Kondo breakdown.
Further theoretical work incorporating interorbital Coulomb terms, multi-impurity interference, or topological aspects is expected to uncover additional nontrivial phases in this emerging realm of heavy-fermion moiré superlattices.

\appendix

\section{Band dispersion of conduction electron}
\label{banddisp}
By disconnecting the coupling $V$ between the $c$- and $f$-subsystems, we can apply the Fourier transformation into momentum space individually for each subsystem. This allows us to diagonalize the bilinear part, leading to the dispersive band $\epsilon_{\vec{k}} + \epsilon^c$ for the $c$-orbitals. However, due to the different lattice spacings of the $c$- and $f$-lattices, their corresponding reduced Brillouin zones, $\text{Bz}^\text{c}$ and $\text{Bz}^\text{f}$, also differ.
Suppressing the spin index $\sigma$ for simplicity, the hybridization term is given by:
\begin{align}
    H^\text{hyb} &= V\sum_l [f^\dagger_l c^{\phantom{\dagger}}_l + \text{h.c.}] \\
                 &= \frac{V}{\sqrt{N_f N_c}} \sum_{\vec{k}\in \text{Bz}^\text{c}} \sum_{\vec{q}\in \text{Bz}^\text{f}} \sum_l [ e^{i (\vec{k}-\vec{q}) \vec{R}_l} f^\dagger_{\vec{q}} c^{\phantom{\dagger}}_{\vec{k}} + \text{h.c.} ] \\
                 &= \frac{V}{\sqrt{N_c^u}} \sum_{\nu=1}^{N_c^u} \sum_{\vec{q}\in \text{Bz}^\text{f}} [f^\dagger_{\vec{q}} c^{\phantom{\dagger}}_{\vec{q}+\vec{p}_{\nu}} + \text{h.c.}]
\end{align}
where $N_c^u = \frac{N_c}{N_f}$ denotes the number of $c$-orbitals within the enlarged unit cell. We use the relation:
\begin{align}
    \sum_l e^{i (\vec{k}-\vec{q}) \vec{R}_l} = N_f\sum_\nu \delta_{\vec{k}, \vec{q}+\vec{p}_{\nu}},
\end{align}
where the set of momentum vectors $\vec{p}_{\nu}$ depends on the basis vectors $\vec{\delta}_i$ of the $f$-lattice, with $\vec{R}_l = \sum_i \alpha_{il}\vec{\delta}_i$. The vectors $\vec{p}_{\nu}$ are determined by the conditions:
\begin{align}
    \label{eq:p_nl}
    \vec{p}_{\nu} \in \text{Bz}^\text{c} \quad \text{and} \quad \vec{p}_{\nu} \cdot \vec{\delta}_i =  2\pi n, \quad n\in \mathbb{N}.
\end{align}
This relation connects $\vec{k} \in \text{Bz}^\text{c}$ to $\vec{q} \in \text{Bz}^\text{f} \subset \text{Bz}^\text{c}$, where $\vec{p}_{\nu}$ are the momentum vectors used to backfold the $c$-band structure into the Brillouin zone $\text{Bz}^\text{f}$.

\begin{figure*}[htb]
\begin{center}
\includegraphics[width=0.99\textwidth]{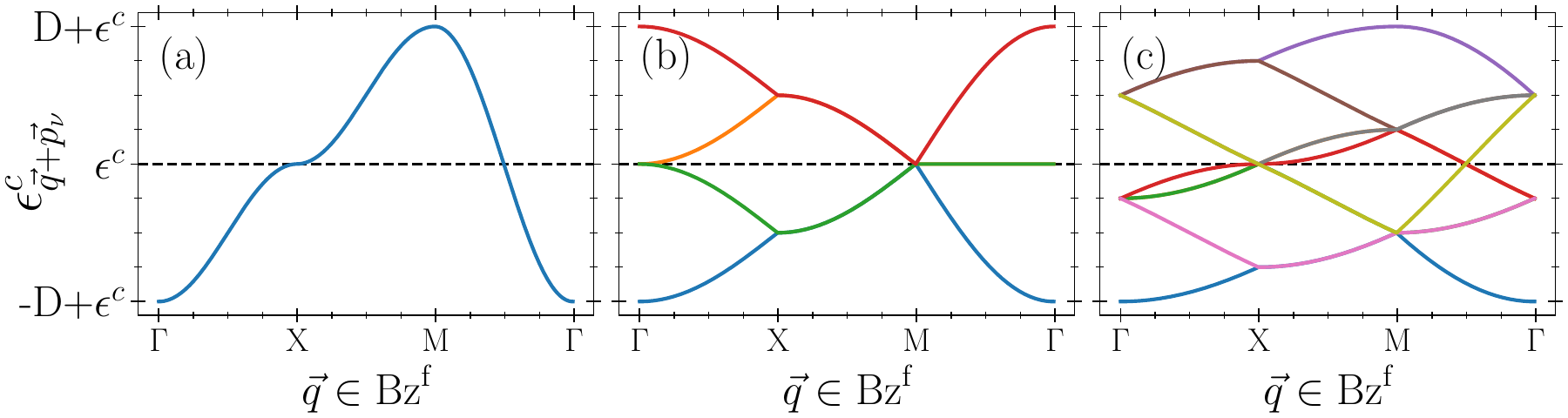}
\end{center}
\caption{Original ($V=0$) $c$-band dispersion of a two dimensional square lattice with nearest neighbor hopping along the high symmetry points $\Gamma$, $\text{X}$ and $\text{M}$, folded back into the reduced Brillouin zone of the $f$-subsystem with $\phi=0$ and (a) $\Delta R^f=1a$, (b) $\Delta R^f=2a$ and (c) $\Delta R^f=3a$. The different colors correspond to different vectors $\vec{p}_\nu$ defined in Eq.~\eqref{eq:p_nl}.} 
\label{fig:A1}
\end{figure*}

Figure~\ref{fig:A1} depicts the original ($V=0$) $c$-band dispersion, $\epsilon_{\vec{k}}$ with $\vec{k}\in \text{Bz}^\text{c}$, of a two dimensional square lattice with nearest neighbor hopping along the high symmetry points $\Gamma$, $\text{X}$ and $\text{M}$, folded back into the reduced Brillouin zone $\vec{q} \in$ Bz$^\text{f}$. Fig.~\ref{fig:A1} (a) corresponds to $\Delta R^f=1a$, (b) to $\Delta R^f=2a$ and (c) to $\Delta R^f=3a$, while $\phi=0$ was maintained for all panels.

\section{Greens functions and Dynamical mean-field theory}
\label{sec:Gf_and-DMFT}

In order to calculate physical properties, we derive the local $c$- and $f$-site Greens functions in this subsection. These Greens functions are also crucial to solve the dynamical mean field self-consistency equation, which we introduce shortly in App.~\ref{sec:DMFT}. It is worth noting that in the following subsections, we derive analytical expressions for real-space $f$- and $c$-site Greens functions that can be evaluated without needing to compute the inverse of a matrix, even for large unit cells.

\subsection{f-site Greens functions}

The single particle Green's function of the $f$-orbitals in the time and frequency domains is defined as
\begin{align}
    \mathcal{G}^f_{\vec{q}}(t) &= \langle\langle f^{\pd}_{\vec{q}}; f^\dagger_{\vec{q}} \rangle\rangle_t = -i \Theta(t) \langle [f^{\pd}_{\vec{q}}(t), f^\dagger_{\vec{q}}(0)] \rangle_0,\\
    \mathcal{G}^f_{\vec{q}}(z) &= \langle\langle f^{\pd}_{\vec{q}}; f^\dagger_{\vec{q}} \rangle\rangle_z = \int_{-\infty}^{\infty} dt \, \mathcal{G}^f_{\vec{q}}(t) e^{i z t},
\end{align}
where $z=\omega + i0^+$ and $\omega \in \mathbb{R}$. Assuming the unit cell is positioned such that the site containing the $f$-orbital is located at $\vec{R} = \vec{0}$, we obtain the local real-space Green's function as
\begin{align}
    \label{eq:Gf_loc}
    \mathcal{G}^f_\text{loc}(\vec{R}=\vec{0},z) = \frac{1}{N_f} \sum_{\vec{q}} \mathcal{G}^f_{\vec{q}}(z).
\end{align}
Using the Dyson equation, we express the full Green's function in terms of the non-interacting part, $\mathcal{G}^\text{0f}_{\vec{q}}(z)$ (which corresponds to $U=0$), and the self-energy $\Sigma_{\vec{q}}(z)$, which accounts for contributions at finite $U>0$:
\begin{align}
    \label{eq:Gf_depleteda}
    \left(\mathcal{G}^f_{\vec{q}}(z)\right)^{-1} = \left(\mathcal{G}^\text{0f}_{\vec{q}}(z)\right)^{-1} - \Sigma_{\vec{q}}(z).
\end{align}
The non-interacting part, \(\mathcal{G}^{0f}_{\vec{q}}(z)\), can be computed by evaluating the equation of motion for the $f$-orbitals:
\begin{align}
    z \langle\langle f^{\pd}_{\vec{q}}; f^\dagger_{\vec{p}} \rangle\rangle_z - \langle\langle [f^{\pd}_{\vec{q}}, H]; f^\dagger_{\vec{p}} \rangle\rangle_z = \delta_{q p},
\end{align}
which yields:
\begin{align}
    \label{eq:Gf_depleted}
    \mathcal{G}^{0f}_{\vec{q}}(z) &= \left( z - \epsilon^f - \Delta_{\vec{q}}(z) \right)^{-1},\\
    \label{eq:Delta_q}
    \Delta_{\vec{q}}(z) &= \frac{V^2}{N_c^u} \sum_{\nu} \mathcal{G}^{0\text{c}}_{\vec{q}, \nu}(z).
\end{align}
Here $\Delta_{\vec{q}}(z)$ is the hybridization function, where  \(\mathcal{G}^{0\text{c}}_{\vec{q}, \nu}(z) = (z - \epsilon_{\vec{q} + \vec{p}_{\nu}})^{-1}\) describes the propagation of an electron in the respective piece \(\nu\) of the non-interacting conduction band, \(\epsilon_{\vec{k}}\), with \(\vec{k} = \vec{q} + \vec{p}_{\nu} \in \text{BZ}^c\) folded into \(\text{BZ}^f\).

\subsection{c-site Greens functions}

In contrast to the \(f\)-orbitals, there are multiple \(c\)-orbitals within the large unit cell, and, consequently, the Greens function has the structure of a matrix. If we concentrate on the diagonal parts in real space, which describe the local properties of the respective \(c\)-orbitals within each unit cell, we need to calculate: 
\begin{align}
    \label{eq:Gc_loc}
    \mathcal{G}^c_{\text{loc}}(\vec{R}_i,z)=\sum_{\vec{q}}\sum_{\mu\nu}e^{i(\vec{p}_\mu-\vec{p}_\nu)\vec{R}_i}\mathcal{G}^c_{\vec{q},\mu\nu}(z),
\end{align}
for the \(c\)-orbital at position \(\vec{R}_i\) relative to the site of the \(f\)-orbital.
\(\mathcal{G}^c_{\vec{q},\mu\nu}(z)=\langle\langle c^{\pd}_{\vec{q}+\vec{p}_\mu}, c^{\d}_{\vec{q}+\vec{p}_\nu} \rangle\rangle_z\) can be derived by evaluating the equation of motion for $U=0$ and replacing $\e^f$ by $\e^f+\Sigma_{\vec{q}}(z)$ to account for interactions on the $f$-orbitals.
This yield:
\begin{align}
\label{eq:Gc_eom}
    \mathcal{G}^c_{\vec{q},\mu\nu}(z)\left(\mathcal{G}^{0c}_{\vec{q},\mu}(z)\right)^{-1} - \sum_{\alpha}\mathcal{G}^c_{\vec{q},\alpha\nu}(z)\frac{V^2}{N^u_c}F_{\vec{q}}(z)=\delta_{\mu\nu},
\end{align}
where we made use of the definition \(F_{\vec{q}}(z)=(z-\epsilon^f-\Sigma_{\vec{q}}(z))^{-1}\). 
From Eq.~\eqref{eq:Gc_eom}, we can now read off the inverse of \(\mathcal{G}^c_{\vec{q},\mu\nu}(z)\), which is given by:
\begin{align}
    \left(\mathcal{G}^{c}_{\vec{q}}\right)^{-1}_{\mu\nu}(z)=\left(\mathcal{G}^{0c}_{\vec{q},\mu}(z)\right)^{-1}\delta_{\mu\nu}-\frac{V^2}{N^u_c}F_{\vec{q}}(z).
\end{align}

As this is a rank-1 update of the diagonal part \(\left(\mathcal{G}^{0c}_{\vec{q},\mu}(z)\right)^{-1}\delta_{\mu\nu}\), we can use the Sherman-Morrison formula \cite{ShermanMorrison} to analytically calculate the inverse:
\begin{align}
\label{eq:Gc_munu}
    \mathcal{G}^c_{\vec{q},\mu\nu}(z) = \mathcal{G}^{0c}_{\vec{q},\mu}(z)\delta_{\mu\nu}+V^2 F_{\vec{q}}(z)
    \frac{\mathcal{G}^{0c}_{\vec{q},\mu}(z)\mathcal{G}^{0c}_{\vec{q},\nu}(z)}
    {1- \frac{V^2}{N^u_c} F_{\vec{q}}(z) \sum_\alpha  \mathcal{G}^{0c}_{\vec{q},\alpha}(z)}.
\end{align}

\subsection{Dynamical mean field theory}
\label{sec:DMFT}

To tackle this complex model in the translational invariant form (PAM), we employ DMFT \cite{DMFT_1, DMFT_2}. By assuming a purely local self-energy, \(\Sigma_{\vec{q}}(z)\approx\Sigma(z)\), we can map the problem onto an effective single-impurity Anderson model. The central DMFT equation therefore reads:

\begin{align}
    \label{eq:DMFT}
    \mathcal{G}^f_{\text{loc}}(z) \stackrel{!}{=} \mathcal{G}^f_{\text{imp}}(z).
\end{align}
Here, \(\mathcal{G}^f_{\text{loc}}(z)\) is defined in Eq.~\eqref{eq:Gf_loc}, and \(\mathcal{G}^f_{\text{imp}}(z)\) describes the local single-particle Greens function of the respective impurity problem:
\begin{align}
\label{eqn:Gimp}
    \mathcal{G}^f_{\text{imp}}(z) &= \left(z-\epsilon^\text{f}-\Delta^\text{eff}(z) - \Sigma(z)\right)^{-1}.
\end{align}
In these equations $\Sigma(z)$ is the correlation-induced part of the self-energy, and $\Delta^\text{eff}(z)$ characterizes the effective medium in which the single impurity is embedded.
From Eq.~\eqref{eq:DMFT} one can formulate the DMFT self consistency equation as
\begin{align}
    \label{eq:DMFT_loop}
    \Delta^\text{eff}(z) = z - \epsilon^\text{f} - \Sigma(z) - (\mathcal{G}^f_\text{loc}(z))^{-1}.
\end{align}
Importantly, $\Sigma(z)$ itself depends on the effective medium $\Delta^\text{eff}(z)$, making Eq. \eqref{eq:DMFT_loop} a self-consistent problem.
The only approximation made by DMFT is to neglect the momentum dependence of the self energy, $\Sigma_{\k}(z)\approx\Sigma(z)$, which gets exact for a lattice with
infinite coordination number, i.e. in infinite dimensions \cite{DMFT_1, DMFT_2}, and large $f$-orbital separations, $\Delta R^f\to\infty$, such that the SIAM limit is reached.

In order to get a self consistent solution we proceed as follows:
\begin{enumerate} 
\item initialize self energy: $\Sigma(z) = 0$
\item calculate $\mathcal{G}^f_\text{loc}(z)$ from Eq.~\eqref{eq:Gf_loc}
\item calculate $\Delta^\text{eff}(z)$ from Eq.~\eqref{eq:DMFT_loop}
\item solve the impurity problem defined by $\Delta^\text{eff}(z)$ to obtain the self-energy $\Sigma(z)$
\item continue steps (II) to (IV) until self-consistency is reached.
\end{enumerate}
It is important to emphasize that the size of the unit cell enters solely through the summation over the back folding vectors $\vec{p}_\nu$ in Eq.~\eqref{eq:Delta_q} when evaluating the lattice $f$-site Green's function in Eq.~\eqref{eq:Gf_loc} for step (2) of the DMFT loop.
Compared with the common method of obtaining $G^f_{\vec{q}}(z)$ from the matrix inversion of the full $(N^u_c+1) \times (N^u_c + 1)$ Green's function, this approach allows us to achieve high numerical accuracy, even when $N^u_c$ is large. Compared with Ref.~\cite{KondoSuperLattice}, we are able to increase the number of Brillouin zone sampling points by a factor of $10^4$, enabling us to resolve very sharp features in the spectral functions for large unit cells.
Moreover, whereas the authors of Ref.~\cite{KondoSuperLattice} were limited to relatively small $f$-orbital separations, $\Delta R^f/a\leq 15$, the improved algorithm allows us to study separations of the order of hundreds of lattice constants.
For the computation of Eq.~\eqref{eq:Gf_loc}, we sample the fourfold reduced Brillouin zone using up to $(10.000 \times 10.000)$ momentum vector points for reaching convergence and perform the summation on an NVIDIA RTX A6000 GPU.

\begin{figure*}[t]
    \centering
    \includegraphics[width=0.99\linewidth]{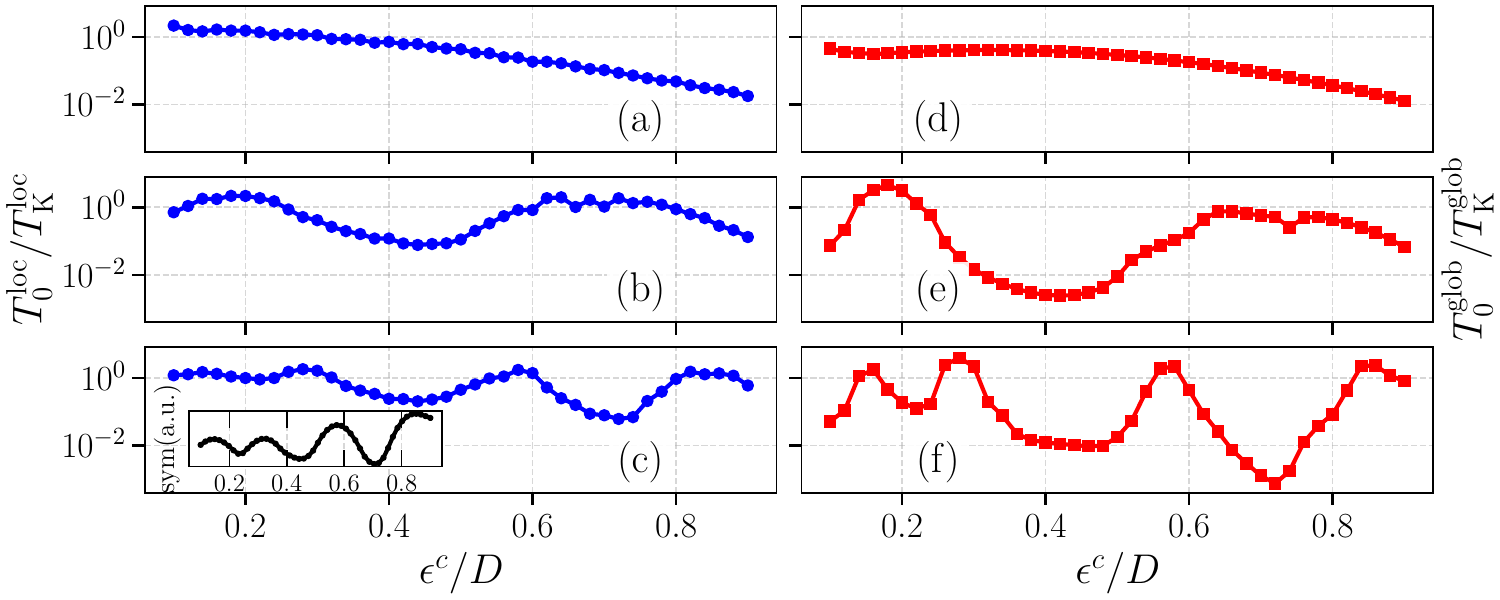}
    \caption{The dependence of low-energy screening scales on the conduction band center $\epsilon^c$ for various $f$-orbital separations and $\phi=0$. The left column [(a), (b), and (c)] illustrates the local scale, while the right column [(d), (e), and (f)] displays the global scale. The $f$-orbital separation increases from top to bottom, corresponding to $\Delta R^f=1a$ in panels (a) and (d), $\Delta R^f=2a$ in panels (b) and (e), and $\Delta R^f=3a$ in panels (c) and (f). Model parameters include $D/\Gamma=5$, $U/\Gamma=10$, and $\epsilon^f=-U/2$. The inset in panel (c) corresponds to the result of evaluating  Eq.~\eqref{eq:sym} in the main text for $\Delta R^f=3a$.
    }
    \label{fig:A2}
\end{figure*}

\section{local and global magnetic moment in NRG calculations}
\label{sec:method_loc_vs_glob}

Within NRG the local moment $\mu^2_\text{loc}$ right at the impurity site and the global magnetic moment $\mu^2_\text{glob}$ corresponding to the whole unit cell can be calculated as follows:
\begin{align}
    \mu^2_\text{loc}(T) &= T\chi_\text{loc}(T) \, ; \quad \chi_\text{loc}(T)=\frac{\partial \langle S_z^f \rangle}{\partial h_z^f}\\
    \mu^2_\text{glob}(T) &= T\chi_\text{glob}(T)\, ; \quad \chi_\text{glob}(T)=\frac{\langle[S^\text{tot}_z]^2\rangle}{T}
\end{align}
Here, $[S^\text{tot}_z]^2$ is a conserved quantum number of the system, so $\langle[S^\text{tot}_z]^2\rangle$ can be directly calculated from the NRG spectrum. $\frac{\partial \langle S_z^f \rangle}{\partial h_z^f}$ is obtained by applying a small magnetic field $h_z$ at the impurity site:
\begin{align}
    \label{eq:H_mod}
    H &\rightarrow \tilde{H} = H + H_\text{mag}^f,\\
    \label{eq:H_mag}
    H_\text{mag}^f &= \frac{h_z}{2}(n^f_\uparrow - n^f_\downarrow),
\end{align}
such that we get $\frac{\partial \langle S_z^f \rangle}{\partial h_z^f} = \lim_{h_z \to 0} \frac{\langle S^f_z\rangle}{h_z}$, where we applied $h_z/V=1e^{-12}$, which is smaller than the lowest accessed temperature scales.
Note that $H^f_\text{mag}$ is only applied to the effective impurity problem after self-consistency has been reached, so we are still considering a paramagnetic system.

\section{Physical mechanism behind the Kondo breakdown in the Lieb-Mattis limit}
\label{sec:mechanism}

When the Brillouin zone $\text{Bz}^f$ of the $f$-subsystem is reduced compared to $\text{Bz}^c$ of the $c$-subsystem, the original ($V=0$) band structure of the $c$-orbitals must be folded back into $\text{Bz}^f$. As a consequence, a single momentum $\mathbf q\in \text{Bz}^f$ labels multiple conduction states $\mathbf q+\mathbf p_\nu$ originating from the unfolded zone, such that the $f$-orbitals are effectively coupled to several backfolded conduction bands, as illustrated in Fig.~\ref{fig:A1} (a)-(c).

In this situation, the local hybridization term couples $f_{\mathbf q}$ coherently to a superposition of backfolded conduction channels with identical amplitude, resulting in an effective rank-one coupling structure. Consequently, the hybridization function governing Kondo screening acquires the form
\begin{equation}
\Delta(\mathbf q,\omega)\propto
\sum_\nu G_c^{0}(\mathbf q+\mathbf p_\nu,\omega),
\label{eq:Delta_interference}
\end{equation}
i.e., a coherent sum over the propagators of the backfolded conduction channels.

The effect of such multi-band extensions of the periodic Anderson model has been studied in detail in Ref.~\cite{MOPAM}, considering the minimal case of two bands. Motivated by the results of Ref.~\cite{MIAM}, it was demonstrated that the low-energy screening scale is determined by the low-frequency properties of the real part of the momentum-dependent hybridization function in Eq.~\eqref{eq:Delta_q}, such that contributions from different bands can interfere destructively. In the extreme case of two bands exactly inverse to each other, these contributions cancel at zero frequency, leading to a breakdown of the Kondo effect \cite{MOPAM} with Power-law spectra and asymptotic $\omega/T$ scaling \cite{OSM_PowerLaw}.

This interference mechanism provides a natural interpretation of the Kondo breakdown observed in the Lieb-Mattis limit of the present model. Whenever the backfolded conduction structure exhibits an approximate mirror symmetry with respect to the Fermi level, channels appear in pairs satisfying
\begin{equation}
\epsilon_{\mathbf q+\mathbf p_\nu}-E_F \approx
-\bigl(\epsilon_{\mathbf q+\mathbf p_{\nu'}}-E_F\bigr),
\end{equation}
which implies opposite low-energy contributions to Eq.~(\ref{eq:Delta_interference}). The resulting cancellation suppresses the effective hybridization at low frequency and thereby reduces the lattice Kondo scale.

For $\Delta R^f=2n\, a$, $\phi=0$ and $\epsilon^c=0$, the original $c$-band folded back into the reduced Brillouin zone $\text{Bz}^f$ of the $f$-subsystem becomes mirror symmetric around the Fermi energy (see Fig.~\ref{fig:A1} (b)). In this perfectly symmetric situation the cancellation is complete, leading to a full extinction of the real part of the momentum-dependent hybridization function at zero frequency,
\begin{equation}
\Re\Delta_{\vec{q}}(0)=0,
\end{equation}
and consequently to the vanishing of the global coherence scale $T^\text{glob}_0$. Weak deviations from this symmetry restore a finite but strongly reduced hybridization, explaining the rapid re-emergence of Kondo screening away from the Lieb-Mattis limit.

An alternative perspective on this mechanism is provided in Appendix~C of Ref.~\cite{OSM_PowerLaw}, where the effect of hybridization is analyzed in terms of a renormalized density of states in combination with analytic predictions from linearized DMFT \cite{Bulla2000}. In that approach, hybridization to additional bands enhances effective tunneling amplitudes and can lead to a divergence of the high-energy contributions to the density of states, suppressing the Mott transition within DMFT. The inclusion of multiple symmetry-related bands restores a finite critical interaction through destructive interference of these hybridization processes. This provides a complementary real-space interpretation of the cancellation mechanism discussed above.

\section{Low energy screening scales with varying conduction band center}
\label{sec:var_mu}

Figure~\ref{fig:A2} depicts the dependency of $T^\text{loc}_0/T^\text{loc}_\text{K}$ [(a), (b), (c)] and $T^\text{glob}_0/T^\text{glob}_\text{K}$ [(d), (e), (f)] on the conduction band center $\epsilon^c/D$.
While keeping the bandwidth and strength of interaction fixed, $D/\Gamma=5$, $U/\Gamma=10$, and $\epsilon^f=-U/2$, the $f$-orbital separation, $\Delta R^f$, increases from top to bottom, while the rotation angle was fixed at $\phi=0$: $\Delta R^f=1a$ in (a) and (d), $\Delta R^f=2a$ in (b) and (e), $\Delta R^f=3a$ in (c) and (f).

In the case of the standard PAM, $\Delta R^f=1a$ [(a), (d)], the dependence of $T^\text{loc}_0$ and $T^\text{glob}_0$ on $\epsilon^c$ is quite featureless.
Starting from $T^\text{loc}_0\approx T^\text{loc}_\text{K}$ and $T^\text{glob}_0\approx T^\text{glob}_\text{K}$ for small $\epsilon^c/D$, both low energy screening scales of the lattice model decrease, reaching a substantially reduced value for large $\epsilon^c/D\to 1$.
This behavior is in agreement with the literature and also occurs in the case of different lattice structures, even when a constant DOS is assumed \cite{PAM_1, PAM_2}.

Increasing the $f$-orbital separation leads to substantial modifications, which are reflected in oscillations of $T^\text{loc}_0$ and $T^\text{glob}_0$ as a function of $\epsilon^c$.
The larger $\Delta R^f$, the more pronounced the oscillations become.

In general, the dependence of $T^\text{loc}_0$ and $T^\text{glob}_0$ on $\epsilon^c/D$ is similar; however, all features are more pronounced in $T^\text{glob}_0$, demonstrating the importance of correlation effects within the non-interacting conduction band, which are induced by the formation of the first-stage Kondo singlet.

The oscillations can be attributed to the effect of destructive hybridization interference \cite{MOPAM}.

As illustrated in the Fig.~\ref{fig:A1} (a)-(c), the original ($V=0$) band structure of the $c$-orbitals requires folding back into the Brillouin zone $\text{Bz}^f$ of the $f$-subsystem for $\Delta R^f>1a$. Consequently, the $f$-orbitals effectively hybridize with multiple bands, which can even intersect or touch each other. By adjusting the band center $\epsilon^c$ to position such crossing points near the Fermi energy, interference effects in the low frequency regime of the real part of the hybridization function become destructive, thereby suppressing overall strength of hybridization. With larger $f$-orbital separations $\Delta R^f$, the presence of more crossing points leads to an increase in the number of oscillations.

To quantify this hypothesis, we measure the degree of mirror symmetry in a small window around the Fermi energy in the original ($V=0$) band structure, back-folded into the Brillouin zone of the $f$-subsystem. This is done by calculating:
\begin{align}
    \label{eq:sym}
    \text{sym} \propto \frac{1}{\sum_{\nu,\vec{q}} g(\epsilon^c_{\vec{q}+\vec{p}_\nu})} \sum_{\vec{q}} \left|\sum_\nu \epsilon^c_{\vec{q}+\vec{p}_\nu} \cdot g(\epsilon^c_{\vec{q}+\vec{p}_\nu})\right|,
\end{align}
where $g(x)$ is a Gaussian function with a width of $0.1 D$ centered around the Fermi energy. A small value corresponds to approximate mirror symmetry, indicating destructive interference in the hybridization function.

The result for $\Delta R^f = 3a$ is depicted in the inset of Fig.~\ref{fig:A2}(c) as a function of the band center $\epsilon^c$, revealing strong oscillations. These oscillations qualitatively match the oscillations of $T^\text{loc}_0$ and $T^\text{glob}_0$, confirming our hypothesis that destructive interference effects within the hybridization function are indeed responsible for the strong dependence of the low-energy screening scales on $\epsilon^c$.

A discussion about a possible connection to oscillations known from the RKKY interaction is given in Appendix~\ref{sec:rkky}.

\section{Connection to the RKKY interaction}
\label{sec:rkky}

In a single-site DMFT treatment, non-local correlations between $f$-orbitals from different unit cells are neglected. However, the oscillatory behavior observed in $T^\text{loc}_0$ and $T^\text{glob}_0$ as $\Delta R^f$ increases prompts consideration of a possible link to the RKKY interaction. The RKKY interaction typically oscillates with frequency $\propto R k_\text{F}$ \cite{RKKY_1, RKKY_2, RKKY_2, TIAM}, where $R$ denotes the distance between two local moments coupled to a conduction band with Fermi wave vector $k_\text{F}$ (where $k_\text{F}$ depends on the band center $\epsilon^c$).

As elucidated in Ref.~\cite{TIAM} in the context of the two-impurity Anderson model (TIAM), the RKKY interaction can be decomposed into two components: one generating the ferromagnetic part and the other responsible for the antiferromagnetic part. It was demonstrated that the latter arises from an effective hopping element $t^\text{eff}(\Delta R^f)$ between the correlated sites, effectively inducing a Heisenberg-type interaction $\propto 4(t^\text{eff})^2/U$ in the strongly interacting limit.

Given that the (depleted) PAM stems from a periodic extension of the TIAM, the individual hopping elements $t^\text{eff}_\text{ij}$ give rise to an effective $f$-band that is crucial when estimating interaction strengths within the system. For instance, in the Lieb-Mattis limit discussed in the main text, the effective hopping elements vanish ($t^\text{eff}_\text{ij}=0$). Consequently, any finite interaction $U/\Gamma>0$ yields strong correlation effects at $T=0$ due to $U/t^\text{eff}_\text{ij}= \infty$. In the absence of PH-symmetry, some $t^\text{eff}_\text{ij}$ always remain finite, effectively limiting $U/t^\text{eff}_\text{ij}<\infty$. Nonetheless, the oscillations observed in $T^\text{loc}_0$ and $T^\text{glob}_0$ still originate from RKKY oscillations of the effective hopping elements and $U/t^\text{eff}_\text{ij}$ respectively.

\normalsize
\clearpage

\vspace{0.5cm}


\begin{acknowledgments}
The work at Los Alamos was carried out under the auspices of the U.S. Department of Energy (DOE) National Nuclear Security Administration under Contract No. 89233218CNA000001. It was support by LANL LDRD Program, and in part by Center for Integrated Nanotechnologies, a DOE BES user facility, in partnership with the LANL Institutional Computing Program for computational resource.
\end{acknowledgments}

\paragraph*{Author contributions}

F.E. conceived the main idea of the project and performed the numerical calculations. J.-X.Z. advised on the motivation of the work and participated in the discussion of results. B.F. devised the project and contributed to the interpretation of the results. All authors participated in the writing and review of the manuscript.

\paragraph*{Data availability}
The data supporting the findings of this study are available from Zenodo \cite{Data} and also on request from the corresponding author.

\bibliography{References}

\end{document}